\documentclass[12pt,preprint]{aastex}
\usepackage{epsf}
\usepackage{times}

\def\reference{\relax\refpar}

\begin{document}

\title{Discovery of a large-scale filament connected to the
massive galaxy cluster MACS\,J0717.5+3745 at $z=0.55$\footnote{Based
in part on data collected at the Subaru Telescope, which is operated by
the National Astronomical Observatory of Japan.}
\footnote{Some of the data presented herein were obtained at the W.M.\
Keck Observatory, which is operated as a scientific partnership among
the California Institute of Technology, the University of California,
and the National Aeronautics and Space Administration. The observatory
was made possible by the generous financial support of the W.M.\ Keck
Foundation.}\footnote{Based partly on observations obtained at the
Gemini Observatory, which is operated by the Association of
Universities for Research in Astronomy, Inc., under a cooperative
agreement with the NSF on behalf of the Gemini partnership: the
National Science Foundation (US), the Particle Physics and
Astronomy Research Council (UK), the National Research
Council (Canada), CONICYT (Chile), the Australian Research Council
(Australia), CNP (Brazil), and CONICET (Argentina).}}

\author{H.\ Ebeling, E.\ Barrett, D.\ Donovan}

\affil{Institute for Astronomy, University of Hawaii, 2680
Woodlawn Drive, Honolulu, HI 96822, USA}

\slugcomment{ApJL, in press}

\begin{abstract}
We report the detection of a 4 $h_{\rm 70}^{-1}$ Mpc long large-scale
filament leading into the massive galaxy cluster MACS\,J0717.5+3745.
The extent of this object well beyond the cluster's nominal virial
radius ($\sim 2.3$ Mpc) rules out prior interaction between its
constituent galaxies and the cluster and makes it a prime candidate
for a genuine filament as opposed to a merger remnant or a double
cluster. The structure was discovered as a pronounced overdensity of
galaxies selected to have $V-R$ colors close to the cluster red
sequence. Extensive spectroscopic follow-up of over 300 of these
galaxies in a region covering the filament and the cluster confirms
that the entire structure is located at the cluster redshift of
$z=0.545$. Featuring galaxy surface densities of typically 15
Mpc$^{-2}$ down to luminosities of 0.10 L$^{\ast}_V$, the most diffuse
parts of the filament are comparable in density to the clumps of red
galaxies found around A851 in the only similar study carried out to
date (Kodama et al.).  Our direct detection of an extended
large-scale filament funneling matter onto a massive, distant cluster
provides a superb target for in-depth studies of the evolution of
galaxies in environments of greatly varying density, and supports the
predictions from theoretical models and numerical simulations of
structure formation in a hierarchical picture.
\end{abstract}


\section{Introduction}

In the now widely accepted hierarchical picture of structure
formation, massive galaxy clusters form through a series of successive
mergers of smaller clusters and infalling groups, as well as through
continuous accretion of matter in their surroundings. Both numerical
simulations (e.g., Colberg et al.\ 2000) and theoretical work (Bond et
al.\ 1996; Yess \& Shandarin 1996) predict that this growth process
proceeds in a highly non-isotropic fashion with infall and mergers
occurring along preferred directions, resulting in spatially highly
correlated structures often referred to as the ``cosmic web'' (Bond et
al.\ 1996).

While there is overwhelming qualitative evidence in support of this
picture from large-scale galaxy redshift surveys (e.g., Geller \&
Huchra 1989; York et al.\ 2000; Colless et al.\ 2001), such surveys
sample only a very small volume to a depth sufficient to probe the
scales of several megaparsecs that characterize the actual infall
region around clusters. Also, galaxies are expected to represent only
a small fraction of the total baryonic mass contained in filaments
which -- according to some estimates -- could rival the mass content
of clusters (Cen \& Ostriker 1999). Observational evidence of diffuse
gas and dark matter in these filamentary structures has, however,
proven difficult to obtain, owing to the low densities involved. The
direct detection of filaments in the vicinity of galaxy clusters thus
remains an observational challenge that is central to our
understanding of the cluster formation process.

In this Letter we present the optical discovery of a large-scale
filament associated with a massive X-ray-selected galaxy cluster at
$z=0.55$.  Throughout we assume a $\Lambda$ cold dark matter cosmology
with $\Omega_M=0.3$, $\Omega_\Lambda=0.7$, and $h_0=H_0/(\mbox{\rm 100
km s$^{-1}$ Mpc$^{-1}$})=0.7$.

\section{Expectations from numerical simulations}

Using simulations conducted by the Virgo consortium, Colberg et al.\
(1999) investigate the relation between the infall pattern around
massive clusters and the large-scale filamentary structure surrounding
them. As expected in a filamentary universe, they find the infall onto
clusters to occur preferentially from distinct directions.
Interestingly, there is a tight correlation between the infall
directions at different times for a given cluster: although different
filaments dominate the infall at different times (redshifts), the
filaments themselves are stable over the full redshift range of the
simulation. While filaments funneling matter onto the cluster
contribute to the observed large-scale structure over a wide range of
scales, all filaments are much more prominent in the vicinity of
massive clusters, the infall region, than in between clusters. Colberg
and co-workers' study also confirms the relevance of filaments for the
baryon budget of the universe: at radii between 4 and 6.5 Mpc more
than 40\% of the total mass is contained in filaments.

\section{Observational evidence of filaments}

Observational evidence of diffuse gas or galaxy overdensities in
three-dimensional space (not to mention dark matter) in filaments is
sparse.  X-ray detections of filaments have been reported for
superclusters and double clusters (Kull \& B\"ohringer 1999, Tittley
\& Henriksen 2001), but the interpretation of these findings as
evidence of filaments has been questioned on the grounds that the
observed emission might be due to gas swept out of these cluster
associations in earlier merger events. Briel \& Henry (1995) find an
upper limit on the X-ray surface brightness of filaments between
clusters of $4\times 10^{-16}$ erg cm$^{-2}$ s$^{-1}$ arcmin$^{-2}$
(0.5--2.0 keV) from {\em ROSAT}\/ All-Sky Survey data. This value is
about an order of magnitude lower than the X-ray surface brightness
detected by e.g., Kull \& B\"ohringer. More recent work by Kodama et
al.\ (2001) and Durret et al.\ (2003) is promising in the sense that
filamentary structures are detected in the vicinity of isolated galaxy
clusters. Unfortunately, the hypothesized filaments all lie well
within the virial radius of the cluster and may thus again be merger
residuals rather than genuine filaments. The most promising possible
detection ($3\sigma$) of a filament thus remains the half-degree long
X-ray feature discussed by Scharf et al.\ (2000) which, at $1.6\times
10^{-16}$ erg cm$^{-2}$ s$^{-1}$ arcmin$^{-2}$ (0.5--2.0 keV), has an
X-ray surface brightness consistent with predictions. Scharf and
coworkers tentatively identify their discovery as a filament extending
over at least 8 Mpc (for an assumed redshift of $z>0.3$). Without
discernible connection to a massive cluster of galaxies within the
study region, this filament is, however, again unlikely to be the kind
of structure predicted to funnel matter onto massive clusters at the
densest vertices of the cosmic web.

\section{MACS\,J0717.5+3745}

The massive galaxy cluster MACS\,J0717.5+3745 [$\alpha$ = 07$^{\rm h}$
17$^{\rm m}$ 31.5$^{\rm s}$, $\delta$ = +37$^{\circ}$ 45' 25''
(J2000)], was independently discovered as the host of a diffuse radio
source by Edge et al.\ (2003) and as a bright X-ray source in the
course of the MAssive Cluster Survey (MACS; Ebeling et al.\
2001). Extensive follow-up observations of MACS\,J0717.5+3745, one of
the most massive clusters known at $z>0.5$ and a severely disturbed
system, have been conducted by the MACS team at energies ranging from
hard X-rays to submillimeter wavelengths. The detection of the
Sunyaev-Zel'dovich effect in the direction of MACS\,J0717.5+3745 is
reported by LaRoque and coworkers (2003); a detailed investigation of
the system's X-ray properties and a weak-lensing study of the dark
matter distribution around MACS\,J0717.5+3745 are in preparation.

Deep imaging observations of MACS\,J0717.5+3745 in the $V$, $R$, $I$,
and $z$ passbands were performed with the SuprimeCam wide-field camera
(Miyazaki et al.\ 2002) on the Subaru 8.3\,m telescope on Mauna Kea in
December 2000 and 2001. SuprimeCam provides a large field of view of
$34'\times27'$ and a 0.2'' pixel$^{-1}$ scale.  The integration times
in $V$, $R$, $I$, and $z$ were $6\times4$, $6\times8$, $7\times6$, and
$9\times3$ minutes, and the seeing (as measured from the final,
co-added frames) was 0.60'', 0.88'', 0.74'', and 0.60'', respectively.

Multi-object spectroscopy (MOS) observations of galaxies in the
MACS\,J0717.5+3745 field were performed on the Keck 10\,m telescope
using the Low Resolution Imaging Spectrograph (Oke et al.\ 1995) in
2000 November, 2002 January, and 2002 November.  Additional MOS
follow-up observations were obtained using the DEep Imaging
Multi-Object Spectrograph on Keck II in 2003 December, and the Gemini
Multi-Object Spectrograph (GMOS; Hook et al.\ 2003) on Gemini North in
2004 March.  Full details of the observational setup are provided by
E.\ Barrett et al.\ (2004, in preparation).  Redshifts were determined
via a multi-template cross-correlation method typically emphasizing
the calcium H and K lines (see Ebeling et al.\ 2004 for details).

\section{Results}

Figure 1 shows a color-magnitude diagram for all galaxies within 2 Mpc
of the nominal cluster center (slightly less than the nominal virial
radius of 2.3 Mpc) as obtained from our Subaru data. The cluster red
sequence is clearly visible. Galaxies within the gray band
(corresponding to $I<22.5$ and $V-R$ values within $\pm 0.2$ mag of
the red sequence) were given highest priority for follow-up
spectroscopy; bluer galaxies were included where permitted by the mask
design constraints.

Figure 2 shows a $20\times 20$ arcmin$^2$ subregion of the $VRz$
SuprimeCam image. Overlaid are the contours of the surface density of
galaxies with $I<22.5$ and $V-R$ colors consistent with the cluster
red sequence (highlighted by the shaded region in Fig.~1).
Adaptive-kernel smoothing has been applied such that the significance
of all features within the bold, red contour is at least 3 $\sigma$.
The full extent of the galaxy overdensity (including the cluster) is
6.3 Mpc, of which about 4.3 Mpc can be attributed to the filament.
This apparently coherent structure thus extends well beyond the virial
radius ($\sim 2.3$ Mpc) of the cluster at its northwestern endpoint,
making it the first convincing candidate for the type of filament
found to channel matter onto massive clusters in the numerical
simulations of, e.g., Colberg and coworkers.  The galaxy surface
densities within the filament are found to range from 10 to 20
galaxies Mpc$^{-2}$ with our $I$ band magnitude limit of 22.5
corresponding to a limiting $V$ band luminosity\footnote{We assume a
value of $M_{\rm V}^{\ast}=-20.23$ (Brown et al.\ 2001) for field
galaxies in the local universe, moderate passive evolution in $M_{\rm
V}^{\ast}$ of $-0.4$ mag out to $z=0.545$, and a typical color of
$V-I=2$ for an elliptical galaxy at $z=0.545$.} of 0.10 L$^{\ast}$ at
$z=0.545$.

To test the hypothesis that the apparently coherent structure evident
in Fig.~2 is indeed a large-scale filament connected to
MACS\,J0717.5+3745, we need to establish whether the color selection
illustrated in Fig.~1 and applied in Fig.~2 selects primarily galaxies
at the cluster redshift.  Figure 3 shows that the spatial distribution
of all galaxies with spectroscopic redshifts as of 2004 March samples
the full extent of the galaxy overdensity of Fig.~2.  Fig.~4
demonstrates (1) that the redshift distribution of galaxies within the
cluster proper is indistinguishable from the one in the filament, and
(2) that our $V-R$ color criterion is efficient at selecting galaxies
at $z\sim 0.55$, i.e., the cluster redshift. We conclude that the
entire filament as shown in Fig.~2 is located at $z=0.55$ and thus at
the same redshift as the massive galaxy cluster at its northwestern
endpoint.

\section{Conclusions}

MACS\,J0717.5+3745 is one of the most massive clusters known at $z>0.5$
with a pronounced double-peaked galaxy distribution indicative of an
ongoing merger event, and a radio relic near the core suggesting
additional recent merger activity (Edge et al.\ 2003). Our discovery
of a large-scale filament leading into the cluster extends the
evidence for significant dynamical activity to scales of at least 5
Mpc from the cluster core, well beyond the virial radius of 2.3
Mpc. Although numerical simulations by Balogh, Navarro \&
Morris (2000) suggest that some fraction of the galaxies observed as
far out as twice the virial radius from the center of a massive
cluster may have been scattered there by previous interactions with
the cluster core, their distribution should be approximately isotropic
and strongly declining with radius. Neither is the case for this
filament which therefore should consist predominantly of matter
approaching the cluster for the first time.

The cluster merger axis points to the north-west in excellent
alignment with the overall orientation of the filament, suggesting that
the current merger activity is due to a group from the filament
falling into the cluster core. The prominence of the filament out to
at least 4 Mpc implies that infall of matter in the same direction
will persist for roughly the next 4 Gyr assuming an infall rate of
$\sim1000$ km s$^{-1}$ ($\sim1$ Mpc Gyr$^{-1}$).

Extending to (and possibly beyond) a distance of 4--5 Mpc from the
cluster core, the filament is detected as a pronounced overdensity of
galaxies with $V-R$ colors within $\pm0.20$ mag of the cluster red
sequence. At measured densities of typically 4 (15, 30) Mpc$^{-2}$ for
red galaxies with $L>0.3$ (0.1, 0.02) $L^{\ast}_V$ at $z=0.545$, the
most diffuse part of the contiguous filament leading into
MACS\,J0717.5+3745 is as dense as the clumps of (mainly) red galaxies
around A851 ($z=0.41$) discovered by Kodama et al.\ (2001). The
prevalence of early-type galaxies in our filament is also in
qualitative agreement with recent studies based on the Two-Degree
Field and Sloan Digitial Sky Survey galaxy catalogs (e.g., G\'{o}mez
et al.\ 2003; Balogh et al.\ 2004), which find the star formation
rates in nearby galaxies ($z<0.1$) to be significantly reduced already
in environments of even lower density.

We conclude that the observed galaxy distribution in and around
MACS\,J0717.5+3745 lends strong support to a picture in which massive
clusters grow via discrete as well as continuous infall of matter
along large-scale filaments. The high galaxy densities observed in the
MACS\,J0717.5+3745 filament and the prevalence of early-type galaxies in
it make this structure a prime target for a detailed investigation of
the physical processes and environmental effects governing the
transition from field to cluster galaxies.  Additional observations
are planned to map the dark matter distribution along the filament via
weak lensing and to measure galaxy morphologies and colors in this
region.

\acknowledgments We thank Ian Smail for allowing us to use some of the
spectroscopic data obtained during his GMOS run of 2004 March 15--17,
as well as for very helpful advice during the preparation of this
Letter. Thanks also to Graham Smith and Alastair Edge for useful
discussions. We are grateful to the University of Hawaii's Telescope
Time Allocation Committee for their support of this study. H.E.\
gratefully acknowledges financial support for MACS from NASA grant NAG
5-8253.

\clearpage

\begin{figure}
\epsffile{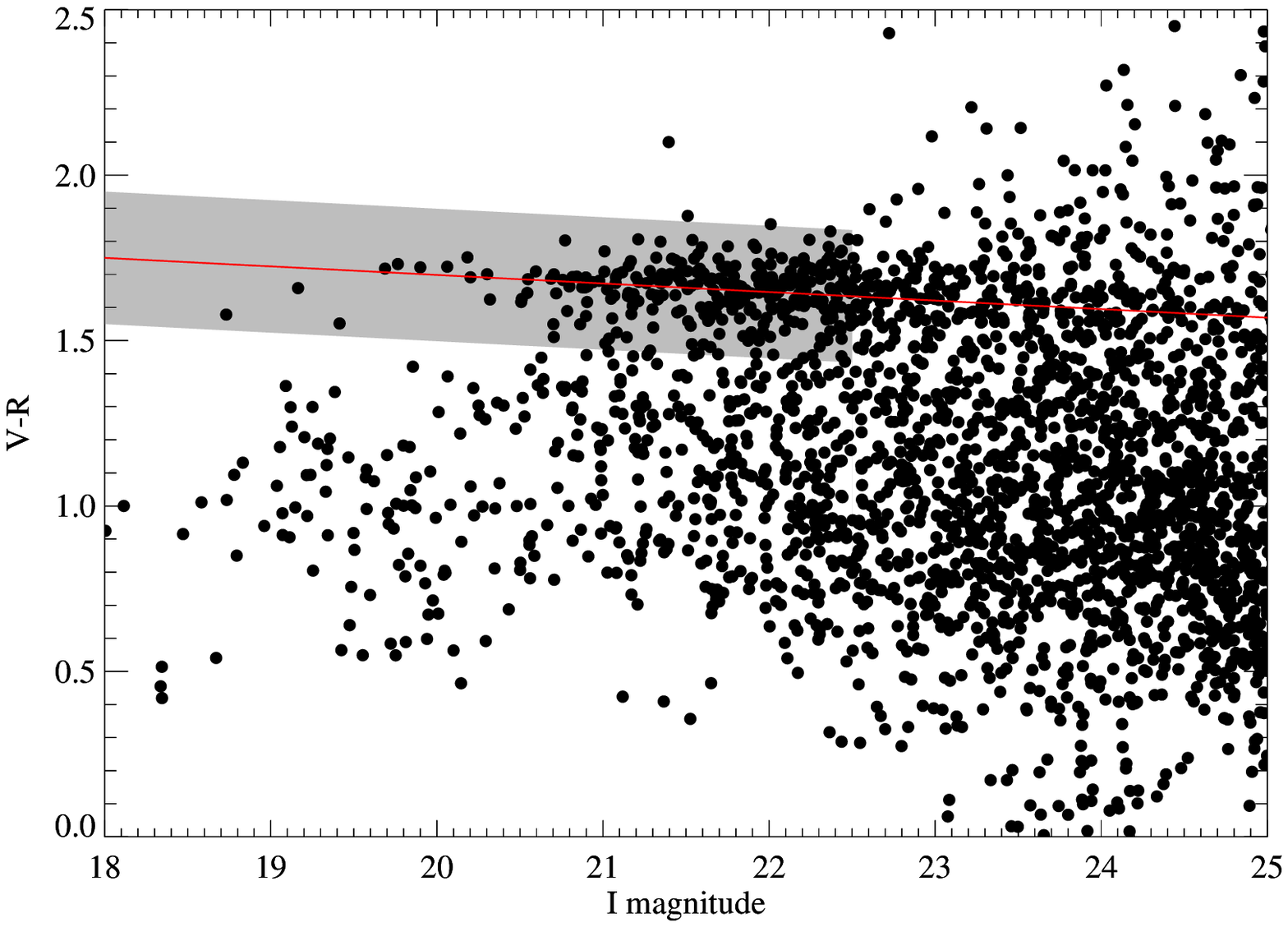}
\caption{Color-magnitude diagram for all galaxies within 2 Mpc of
the nominal cluster position. The red line marks the cluster red
sequence; galaxies selected at highest priority for spectroscopic
follow-up observations fall within the shaded band around the red sequence.}
\end{figure}

\clearpage

\begin{figure}
{\LARGE See separate file 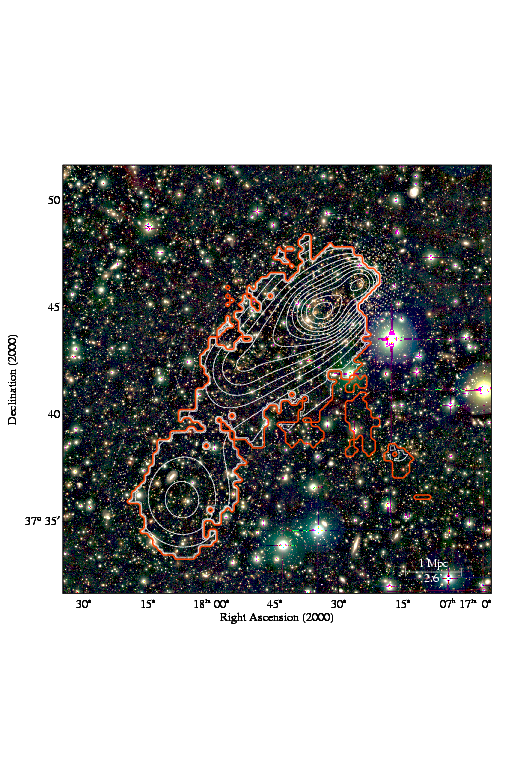}\\
\caption{$VRz$ color image of a $20\times 20$ arcmin$^2$ region around
MACSJ0717.5+3745 as obtained with SuprimeCam. The overlaid contours
show the adaptively smoothed surface density distribution of galaxies
with $I<22.5$ and $V-R$ colors following the cluster red sequence
(cf.\ the shaded region in Fig.~1). Contours are spaced
logarithmically, with the lowest contour corresponding to 10 galaxies
Mpc$^{-2}$ (50\% above the background value), and the levels of
adjacent contours differing by 20\%. Contours are shown dotted outside
the region within which the adaptive smoothing criterion (significance
greater than $3\sigma$) is met ({\em bold red contour}). The contour
marking the shallow peak of the south-eastern end of the filament
marks a local peak galaxy surface density of 20.7 galaxies
Mpc$^{-2}$. Since the bottom edge of the shown region is also the edge
of the overall Subaru SuprimeCam image, it is conceivable that the
filament in fact extends yet farther south than suggested by the
galaxy surface density contours shown here.}
\end{figure}

\clearpage

\begin{figure}
\epsffile{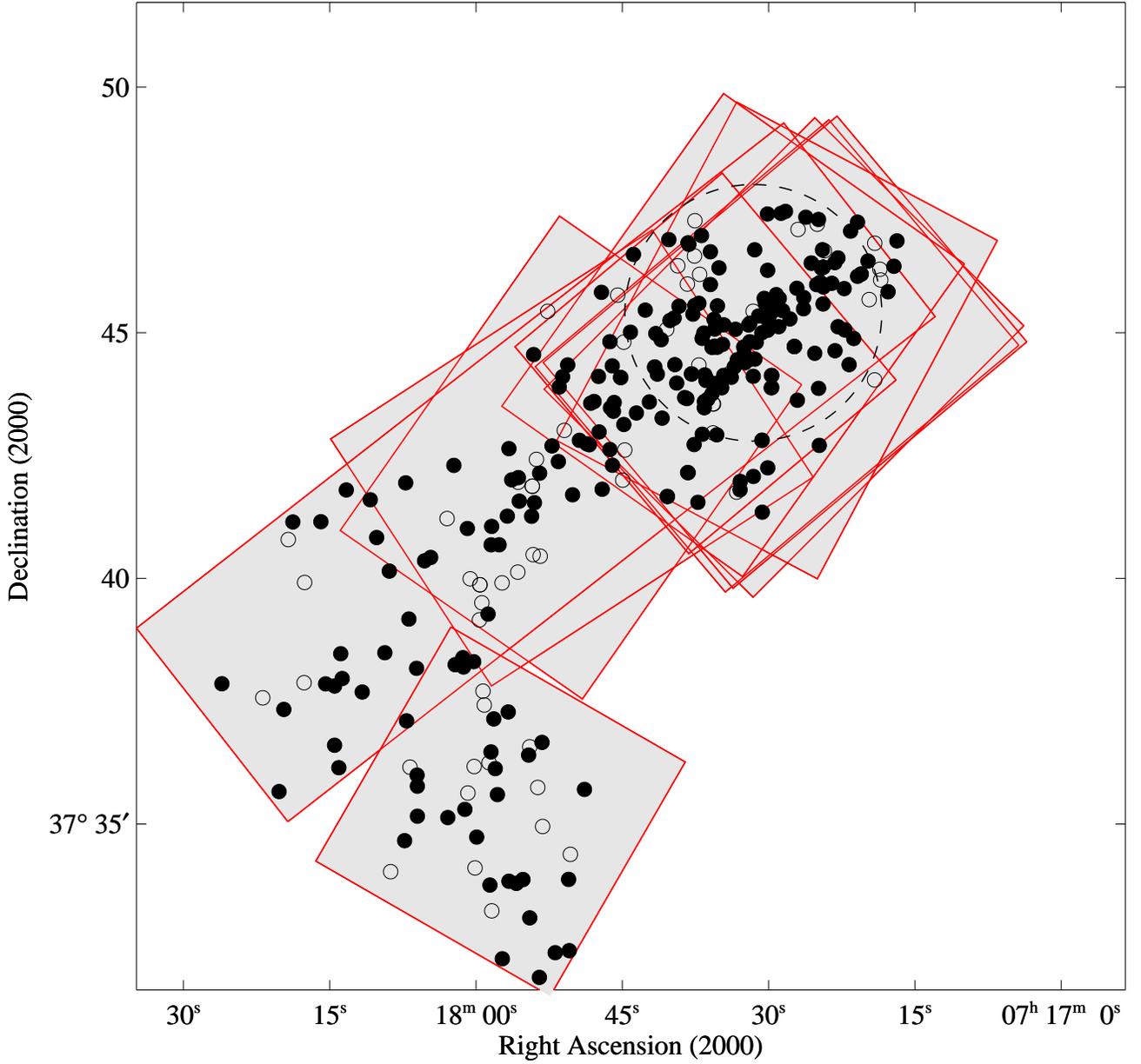}
\caption{Locations of all galaxies with spectroscopic redshifts as of
2004 March; filled symbols mark galaxies with redshifts between 0.52
and 0.57, i.e., the 3 $\sigma$ range around the systemic redshift of
MACS\,J0717.5+3745 (see also Fig.~4). The area shown is the same as
depicted in Fig.~2. The shaded regions and box outlines mark the
fields targeted with the individual MOS masks. The dotted line marks a
circle of 1 Mpc radius around the nominal cluster center.}
\end{figure}

\clearpage

\begin{figure}
\epsffile{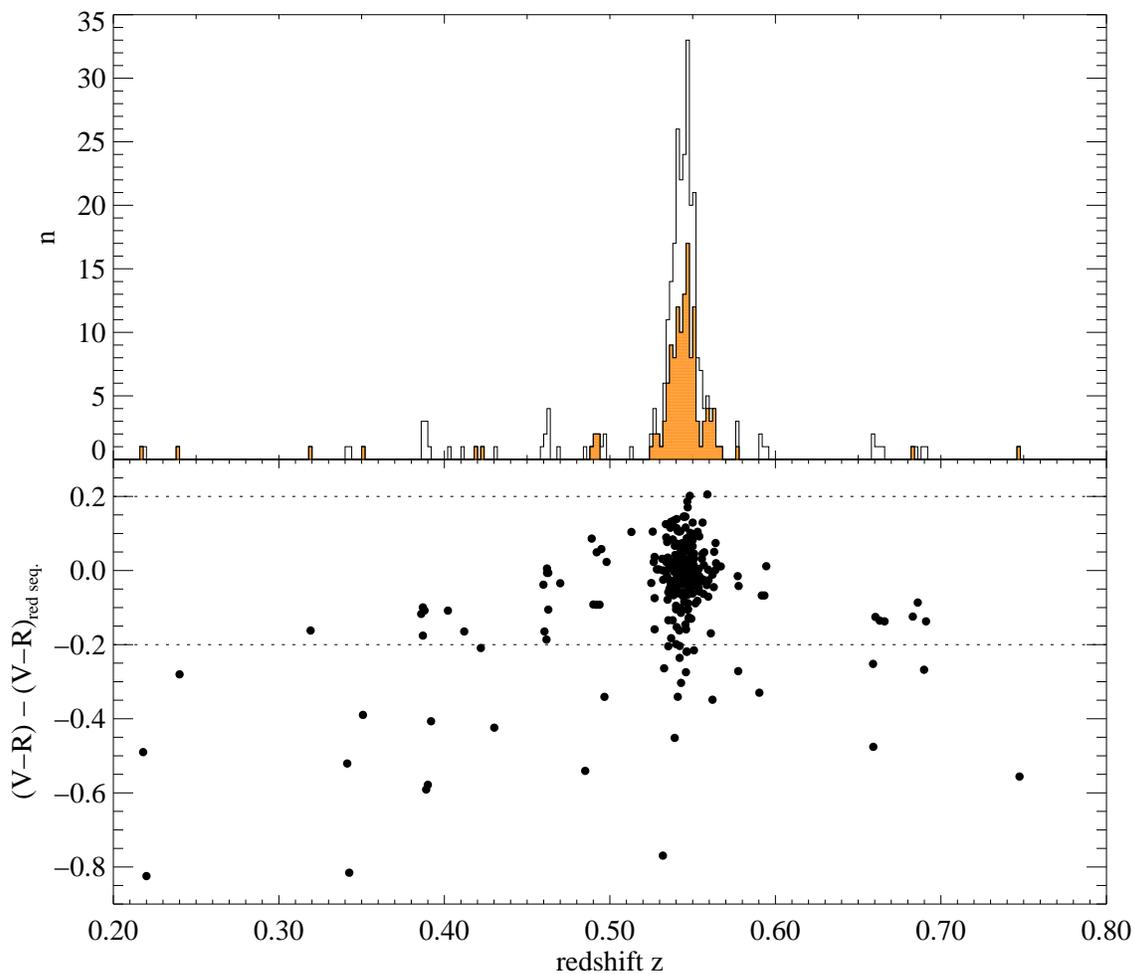}
\caption{{\em Top}: Histogram of all 302 redshifts obtained by us so
far in the regions highlighted in Fig.~3. The shaded histogram
represents the redshift distribution of galaxies within 1 Mpc of the
cluster core ({\em Fig.~3, dashed circle}). {\em Bottom}: Galaxy
redshift vs.\ offset in $V-R$ color from the cluster red sequence for
all galaxies observed so far. Our color criterion efficiently selects
galaxies at and around $z=0.55$.}
\end{figure}

\end{document}